`

# Giant Uniaxial Magnetocrystalline Anisotropy in $SmCrGe_3$

Mingyu Xu,[1] Yongbin Lee[2], Xianglin Ke,[3] Min-Chul Kang,[2] Matt Boswell,[1] Sergey. L. Bud'ko,[2, 4] Lin Zhou,[2, 5] Liqin Ke[2], Mingda Li,[6,7] Paul. C. Canfield,[2, 4*] Weiwei Xie[1*]

[1]Department of Chemistry, Michigan State University, East Lansing, Michigan 48824, USA
[2]Ames National Laboratory, Iowa State University, Ames, Iowa 50011, USA
[3]Department of Physics and Astronomy, Michigan State University, East Lansing, Michigan 48824, USA
[4]Department of Physics and Astronomy, Iowa State University, Ames, Iowa 50011, USA
[5]Department of Materials Science and Engineering, Iowa State University, Ames, IA 50011, USA
[6]Quantum Measurement Group, MIT, Cambridge, MA 02139, USA
[7]Department of Nuclear Science and Engineering, MIT, Cambridge, MA, 02139

## *Abstract*

Magnetic anisotropy is a crucial characteristic for enhancing spintronic device performance. The synthesis of $SmCrGe_3$ single crystals through a high-temperature solution method has led to the determination of uniaxial magnetocrystalline anisotropy. Phase verification was achieved using scanning transmission electron microscopy (STEM), powder, and single-crystal X-ray diffraction techniques. Electrical transport and specific heat measurements indicate a Curie temperature ($T_C$) of approximately 160 K, while magnetization measurements were utilized to determine the anisotropy fields and constants. Curie-Weiss fitting applied to magnetization data suggests the contribution of both Sm and Cr in the paramagnetic phase. Additionally, density functional theory (DFT) calculations explored the electronic structures and magnetic properties of $SmCrGe_3$, revealing a significant easy-axis single-ion Sm magnetocrystalline anisotropy of 16 meV/f.u.. Based on the magnetization measurements, easy-axis magnetocrystalline anisotropy at 20 K is 13 meV/f.u..

`

## Introduction

Magnetic anisotropy (MA) is a pivotal property of materials that significantly influences the performance of modern spintronic devices [1]. The interest in MA is attributed to its critical role in various physical phenomena, including the permanent and topological magnets, single-molecule magnets, the Kondo effect, magnetocaloric effects, the magnetic skyrmion dynamics, and so on [2–7]. Therefore, the large magnetic anisotropy and coercivity are the key properties of functional magnetic materials. From an engineering perspective, though coercivity can be modified by geometrical shaping, [8] the anisotropy field sets the upper limit for the coercivity. In contrast, from a materials chemistry standpoint, the intrinsic magnetocrystalline anisotropy presents a promising avenue for MA control. This intrinsic anisotropy is primarily influenced by the local crystal symmetry and the spin-orbit coupling (SOC) associated with the magnetic ions.

Our magnetism research is concentrated on synthesizing novel magnetic materials through the combination of 3*d* transition-metal and 4*f* rare earth elements, which usually display a range of tunable magnetic functionalities. In many cases of RE-TM compounds, both 3*d* and 4*f* electrons present magnetic moment and contribute to interesting magnetic proprieties, such as the high Curie temperature ($T_C$), and large anisotropy, as in 3*d* compounds which usually have the high $T_C$, and the 4*f* compounds most of which have high magnetocrystalline anisotropy. For example, compounds from the Sm-Co family [9] and $Nd_2Fe_{14}B$ [9–11] are prevalent choices within industrial applications due to their substantial contributions to magnetic performance. $SmCo_5$ is renowned for its significant magnetocrystalline anisotropy and high coercivity field [1,10]. In general, then, to achieve the industrial applications of magnets, both 3*d* and 4*f* electrons are needed to offer large magnetocrystalline anisotropy and high operation temperature.

*RE*$CrGe_3$ compounds, with *RE* representing La, Ce, Pr, Nd, and Sm, crystallize in a hexagonal structure and exhibit intricate magnetic properties due to their unique geometric arrangement of $CrGe_6$ clusters and the spatial positioning of Cr and *RE* atoms [11–16]. $LaCrGe_3$ has garnered extensive investigation owing to its nuanced ferromagnetic properties by varying pressures and distinct magnetic domain behaviors induced by high-field

`

quenching [12,13,17–22]. Samarium compounds, in particular, display more complex magnetic characteristics due to closely spaced multiplet levels [23] and the significant orbital contribution of $Sm^{3+}$ ions, which enriches the study of magnetocrystalline anisotropy in permanent magnets anisotropy, necessitating synthesizing its single-crystalline form for detailed analysis.

In this study, we report the synthesis of single-crystalline $SmCrGe_3$ and its phase and structural characterization using a comprehensive suite of analytical techniques, including scanning transmission electron microscopy (STEM), energy dispersive spectroscopy (EDS), powder, and single-crystal X-ray diffraction. By conducting magnetization measurements at various temperatures and different magnetic field orientations in conjunction with theoretical calculation, we showed that $SmCrGe_3$ possesses a giant easy-axis magnetocrystalline anisotropy, reaching 13 meV/f.u. at 20 K.

## Experimental Parts and Calculations

### Crystal Growth, Structural Characterization, and Magnetic Measurements

Single crystals of $SmCrGe_3$ were synthesized at Ames National Laboratory utilizing a high-temperature solution growth methodology, a technique delineated in references [13,24]. This synthesis process entailed a two-step approach. Initially, a mixture consisting of Sm pieces (SM-TWE-0001AM, Ames National Laboratory), Cr pieces (99.996%, Alfa Aesar), and Ge pieces (99.999%, MSE Supplies) with an atomic ratio of 18:12:70 was placed into a Canfield Crucible Set (CCS) [25,26] and subjected to a thermal regimen where the temperature was heated up to 1180 °C. Subsequently, the system underwent a controlled cooling process to 825 °C for 20 hours. At 825 °C, a mixture of phases was separated from the liquid by a lab-centrifuge. Secondly, the decanted liquid was resealed, heated to 850 °C (remelt it fully), and slowly cooled from 850 °C to 810 °C over roughly 15 hours. At 810 °C, the growth was decanted, and the $SmCrGe_3$ crystalline solid phase was separated from excess liquid.

`

To show the crystalline structure and identify potential defects within the $SmCrGe_3$ single crystals, a specimen with dimensions of 0.092 × 0.064 × 0.048 $mm^3$ was selected for analysis. This crystal was affixed to a nylon loop using Paratone oil, facilitating its examination via a Rigaku XtalLAB Synergy, Dualflex, Hypix single crystal X-ray diffractometer. The apparatus was operated at room temperature. Crystallographic data acquisition was conducted employing $\omega$ scan methodology, utilizing Mo Kα radiation (λ = 0.71073 Å) emitted from a micro-focus sealed X-ray tube under operating conditions of 50 kV and 1 mA. The determination of the experimental parameters, including the total number of runs and images, was derived algorithmically from the strategy computations facilitated by the CrysAlisPro software, version 1.171.42.101a (Rigaku OD, 2023). Subsequent data reduction processes incorporated corrections for Lorentz and polarization effects. Integration of the collected data, predicated on a hexagonal unit cell model, yielded a dataset comprising 5510 reflections up to a maximum $2\theta$ angle of 82.192. Of these reflections, 262 were identified as independent, achieving an average redundancy of 20, with a completeness of 100% and a $R_{int}$ value of 7.64%. An advanced numerical absorption correction was implemented, leveraging Gaussian integration across a model of a multifaceted crystal [27]. Moreover, an empirical absorption correction employing spherical harmonics was applied within the SCALE3 ABSPACK scaling algorithm to refine the data further [28]. **Tables S1** and **S2** show the results of the single-crystal XRD. The structure was solved and refined using the Bruker SHELXTL Software Package [29,30], using the space group $P6_3/mmc$, with Z = 2 for the formula unit, $SmCr_{0.906(9)}Ge_3$. The final anisotropic full-matrix least-squares refinement on $F^2$ with 11 variables converged at $R_1$ = 2.60 %, for the observed data and $wR_2$ = 6.72 % for all data. The goodness-of-fit was 1.115. The largest peak in the final difference electron density synthesis was 3.87 $e^-/Å^3$, and the largest hole was -2.60 $e^-/Å^3$ with an RMS deviation of 0.402 $e^-/Å^3$. Based on the final model, the calculated density was 7.576 $g/cm^3$ and F (000), 359 $e^-$.

Powder X-ray diffraction (PXRD) analysis was also performed. The $SmCrGe_3$ crystals were grounded using an agate mortar and pestle to achieve a homogenous powder. This powdered sample was then uniformly distributed on a single crystalline silicon sample holder, designed for zero background measurements, with a minimal application of vacuum grease to secure the powder in place. The PXRD data acquisition at room temperature spanned a 2θ range from

`

15° to 100°, utilizing incremental steps of 0.01° and a fixed dwell time of 3 seconds per step. These measurements were conducted using a Rigaku MiniFlex II powder diffractometer, employing Bragg-Brentano geometry coupled with Cu Kα radiation (λ = 1.5406 Å). The refinement of the powder X-ray data was executed using the GSAS-II software suite [31], and the occupancy of Cr is 0.931(7).

The phase composition was analyzed employing a JEOL 6610LV scanning electron microscope equipped with a tungsten hairpin emitter (JEOL Ltd., Tokyo, Japan). For elemental analysis, energy-dispersive X-ray spectroscopy (EDX) was conducted utilizing an Oxford Instruments AZtec system (Oxford Instruments, High Wycomb, Buckinghamshire, England), operating software version 3.1. This setup included a 20 $mm^2$ Silicon Drift Detector (SDD) and an ultra-thin window integrated with the JEOL 6610LV SEM. The single crystals of $SmCrGe_3$ were affixed to carbon adhesive tape and introduced into the SEM chamber for examination at an accelerating voltage of 20 kV. Data acquisition entailed collecting spectra at multiple points along the individual crystals over an optimized timeframe. Quantitative compositional analysis was performed using SEM Quant software, which applies corrections for matrix effects to the intensity measurements. The occupancy of Cr is 0.93(2) given by the results of EDS, in agreement with the XRD results discussed above.

Transmission electron microscope (TEM) samples were prepared by a focused ion beam instrument with a gas injection system (Helios, Thermo Fisher Scientific Ltd.). At room temperature, the TEM samples were investigated using an aberration-corrected TEM (Titan Cube, Thermo Fisher Scientific Ltd.) at 200 kV.

Temperature- and magnetic-field-dependent DC and VSM magnetization and resistance measurements, as well as temperature-dependent specific heat measurements, were carried out using Quantum Design (QD), Magnetic Property Measurement Systems (MPMS3), and Physical Property Measurement System (DynaCool). Temperature- and field-dependent DC and VSM magnetization measurements were taken for *H* parallel and perpendicular to the

`

crystallographic $c$-axis by placing the rod-like sample between two collapsed plastic straws with the third, uncollapsed, straw providing support as a sheath on the outside or by using a quartz sample holder. Samples were fixed on the straw or quartz sample holder by GE-7031-varnish. In the VSM magnetization assessments, an oscillation peak amplitude of 4 mm and a mean acquisition time of 2 seconds were employed to ensure precise data collection. The demagnetizing factor, $N$, is estimated by the dimensions of the sample [32]. $N < 0.1$ as the field is parallel to the $c$-axis, and $N < 0.5$ as the field is perpendicular to the $c$-axis. DC electrical resistance measurements were performed in a standard four-contact geometry using the ACT option of the PPMS. 50 µm diameter Pt wires were bonded to the samples with silver paint (DuPont 4929N) with contact resistance values of about 2-3 Ohms. Specific heat capacity measurements under varying temperature conditions were executed using the relaxation method.

**Theoretical Simulation on Magnetocrystalline Anisotropy**

The density functional theory (DFT) calculations are carried out to investigate the band structure and intrinsic magnetic properties of $SmCrGe_3$. Although Cr vacancies are present in the real samples, our calculations focus on the stoichiometric structure. Experimental lattice parameters and atomic coordinates (Tables S1 and S2) are adopted for the calculations. The calculations are performed using a full-potential linear augmented plane wave (FP-LAPW) method, as implemented in Wien2K [33]. The generalized gradient approximation of Perdew, Burke, and Ernzerhof [34] is used for the correlation and exchange potentials. Spin-orbit coupling (SOC) is included using a second variational method. To generate the self-consistent potential and charge, we employed $R_{MT} \cdot K_{max} = 8.0$ with muffin-tin (MT) radii $R_{MT} = 2.8$, 2.2 and 2.2 a.u. for Sm, Cr, and Ge atoms, respectively. The calculations are performed with 1224 k-points in the irreducible Brillouin zone (IBZ) and iterated until the charge differences between consecutive iterations are smaller than $10^{-5}$ e and the total energy differences lower than $10^{-3}$ mRy/cell. For the density of states (DOS) calculation, a denser k-mesh with 1710 k-points in the IBZ is used. The strongly correlated Sm-4f electrons are treated using the DFT+U method with the fully-localized-limit (FLL) double-counting scheme. For magnetic properties calculations using DFT-based methods, the orbital dependence of self-interaction error can often contradict

`

Hund’s rules and plague MA calculations. Therefore, the 4f configurations of the converged solutions should be carefully monitored to avoid unphysical results. Detailed discussions of challenges and methods of MA calculations can be found in [35].

## Results and Discussion

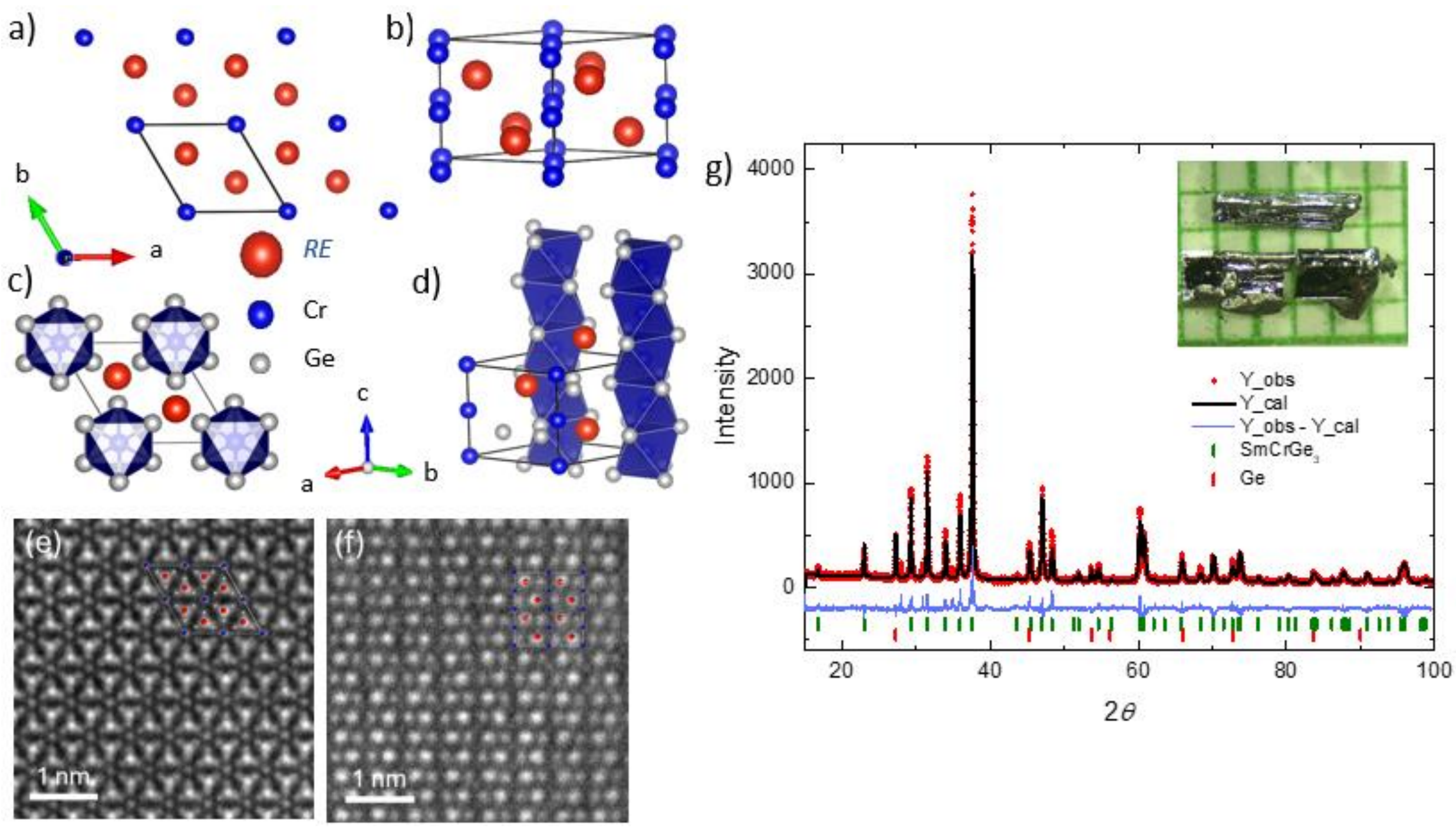

**Fig. 1. Single crystal structure and powder X-ray diffraction pattern of $SmCrGe_3$.** (***a***, ***c***) structural viewpoint and (***e***) high angle annular dark field (HAADF) STEM image from the *c*-axis; (***b***, ***d***) structural viewpoint and (***f***) HAADF STEM image from ab-plane. (***g***) Power X-ray diffraction data at room temperature with the Rietveld refinement done by GSAS-II shown with the prime phase being $SmCr_{0.93}Ge_3$ and the minor phase Ge. The red dots indicate the intensity measured, the black line shows the fitting, and the blue line presents the residuals of the fitting. **(Inset)** The crystal picture is over a millimeter paper grid.

**Hexagonal Perovskite Structure and Defects in $SmCrGe_3$:** Utilizing single crystal X-ray diffraction (SXRD) analysis, we confirm that $SmCrGe_3$ crystallizes in a known hexagonal close-packed structure characterized by alternating layers of $SmGe_3$ and $CrGe_6$ octahedra. [11] These $CrGe_6$ octahedra are uniquely arranged in columns, sharing a pair of opposing faces, forming linear chains of chromium (Cr) atoms along the crystallographic c-axis. The interatomic Cr-Cr distance within these chains is measured at 2.83 Å, noticeably shorter than the Cr-Cr separation observed in elemental chromium (2.89 Å). This structural motif is encapsulated within the space group $P6_3/mmc$ (No. 194), as shown in the schematic diagram of **Fig. 1*a*-1*d*** and the corresponding high-angle annular dark field STEM image of **Fig.1*e*** and **1*f***. $ABX_3$ compounds, where A denotes a large cation, B a small cation, and X an anion, are broadly categorized into various structural types based on the stacking sequences of their $AX_3$ layers, known as perovskite structures. These include configurations with two-layer, three-layer, and six-layer arrangements, among others. In the cubic close-packed variation of these structures, $BX_6$ octahedra interconnect exclusively via their vertices. Contrastingly, $SmCrGe_3$ exhibits hexagonal close-packing of $SmGe_3$ layers, with $CrGe_6$ octahedra columnar stacking through face-sharing interactions, as previously mentioned. A comparative analysis reveals that, from a purely ionic bonding perspective, $SmCrGe_3$'s structural configuration is ostensibly less stable than that of the cubic perovskites due to the significant interionic repulsion prompted by the shortened Cr-Cr distances. It is noted that larger A-site cations (e.g., Ba, La) can mitigate this repulsion by stabilizing the hexagonal close-packing arrangement of the $AX_3$ layers, thus favoring the face-sharing geometry of the $BX_6$ octahedra. In the context of $SmCrGe_3$, the ionic radius of $Sm^{3+}$ (109 ppm) is comparatively smaller than that of $La^{3+}$ (117 ppm), raising inquiries regarding the mechanism by which $Sm^{3+}$ accommodates such compact Cr-Cr distances. Indeed, SXRD analyses reveal approximately 10% vacancies at the Cr sites, a defect structure that likely ameliorates the otherwise pronounced Cr-Cr repulsive interactions. **Fig. 1*g*** presents the Rietveld refinement of powder X-ray diffraction data, indicating the presence of solely $SmCrGe_3$ and minor Ge flux, with an inset displaying a photograph of the single-crystalline specimens, characterized by their metallic luster and bar-like morphology. The details of single-crystal X-ray diffraction measurements are shown in **Table S1** and **S2**. The occupancy of Cr is 0.906(9). Compared with occupancy 0.931(7) from power X-ray diffraction and 0.93(2) for EDS (**Fig. S1**), the occupancy number from single crystal X-ray diffraction is larger than PXRD, which may be due to the measurement conditions and sample variation. However, these data are consistent with EDS results and give more than 5% vacancy in the Cr site. Instead of using $SmCr_{0.906}Ge_3$ or $SmCr_{0.931}Ge_3$, we use $SmCrGe_3$ to denote the compound throughout the text.

`

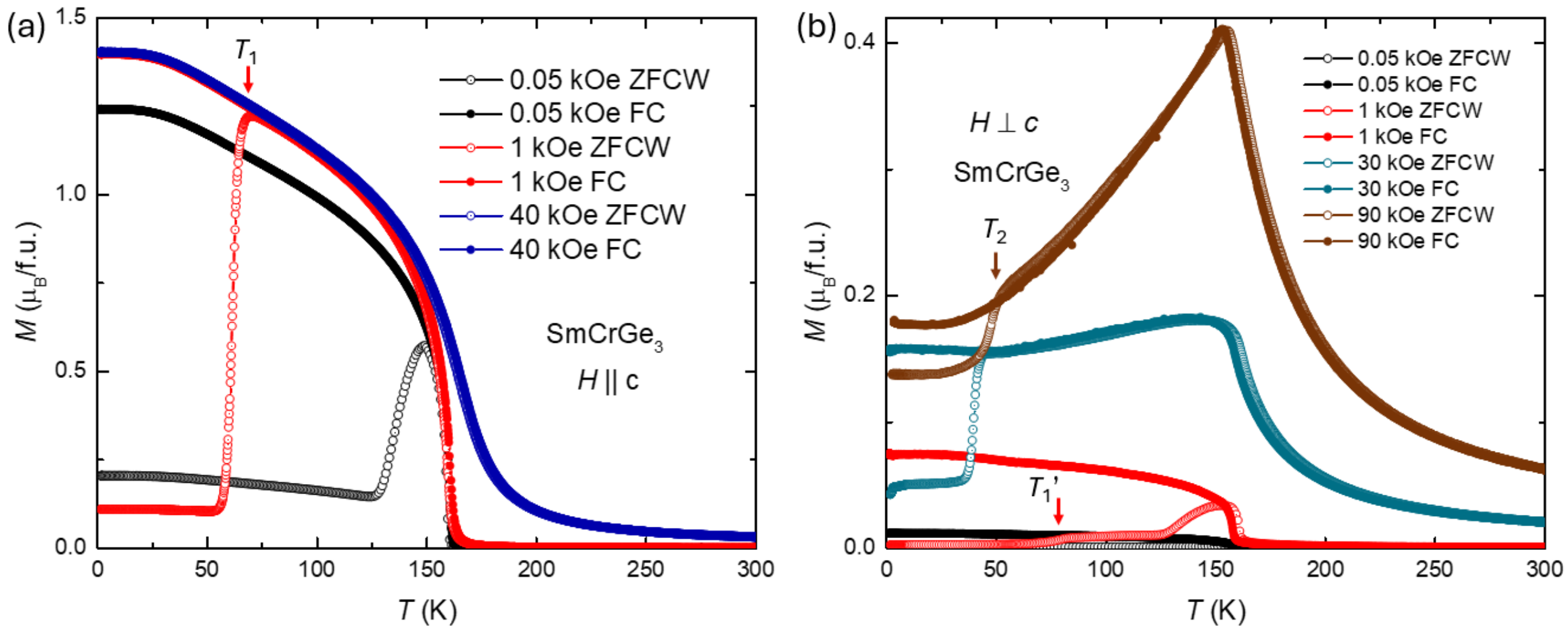

**Fig. 2. Temperature-dependent magnetization of $SmCrGe_3$ measured along different directions at various applied fields in zero-field-cooled-warming (ZFCW) and field-cooled-cooling (FCC) temperature protocols. Fig. 2*a*** shows the measurements taken along the *c*-axis. $T_1$ is the feature temperature that characterizes the jump-like feature as the field is 1 kOe. **Fig. 2*b*** gives the measurements taken perpendicular to the *c*-axis. $T_1'$ and $T_2$ are the feature temperatures that characterize the kink-like and jump features.

**Fig. 2** presents the analysis of temperature-dependent zero-field-cooled-warming (ZFCW) and field-cooled-cooling (FCC) magnetization measured under various magnetic fields applied parallel or perpendicular to the *c*-axis. **Fig.S2** presents the $M(T)$ measurements under more magnetic fields. The ferromagnetic transition is around 159.6 K, which is also evident via the zero-field temperature-dependent electrical transport and specific heat measurements in **Fig.S4**. **Fig.2*a*** shows the temperature-dependent magnetization as the field is applied parallel to the *c*-axis, which presents the irreversibility at a low temperature. The temperature of the jump-like features due to the domain elimination during warming decreases as the magnetic field increases. No hysteresis was observed in **Fig.2S** as the magnetic field above 30 kOe. As shown in **Fig.2*b***, when the magnetic field is applied in the direction perpendicular to the *c*-axis, several $M(T)$ differences are observed from the other direction. First, the maximum moment, ~0.4 $\mu_B$ (90 kOe), is much smaller than 1.4 μB (30 kOe) in $H \parallel c$. Second, the jump features are also observed, but these features still exist in high magnetic fields, which may be due to a few degrees of sample misalignment relative to the magnetic field direction. The kink-like features, shown as $T_1'$, appear at similar temperatures as $T_1$. Finally, the decrease of the moment as the temperature decreases appears after the transition with the high magnetic

`

field in $H \perp c$. There is no confirmed explanation for the moment decrease under the magnetic field larger than 16 kOe, as shown in **Fig.2S*b*** and **2S*d***. There is one suspicion for magnetic moment decrease. This decrease in the non-domain-motion range (the range ZFCW and FCC overlap) indicates no domain change. The magnetization decrease means that the spins move away from the magnetic field direction and rotate to the easy direction. This is the competition between the Zeeman interaction and the magnetocrystalline anisotropy. The different magnetization responses to magnetic fields applied parallel versus perpendicular to the *c*-axis further accentuate the anisotropic nature of the material.

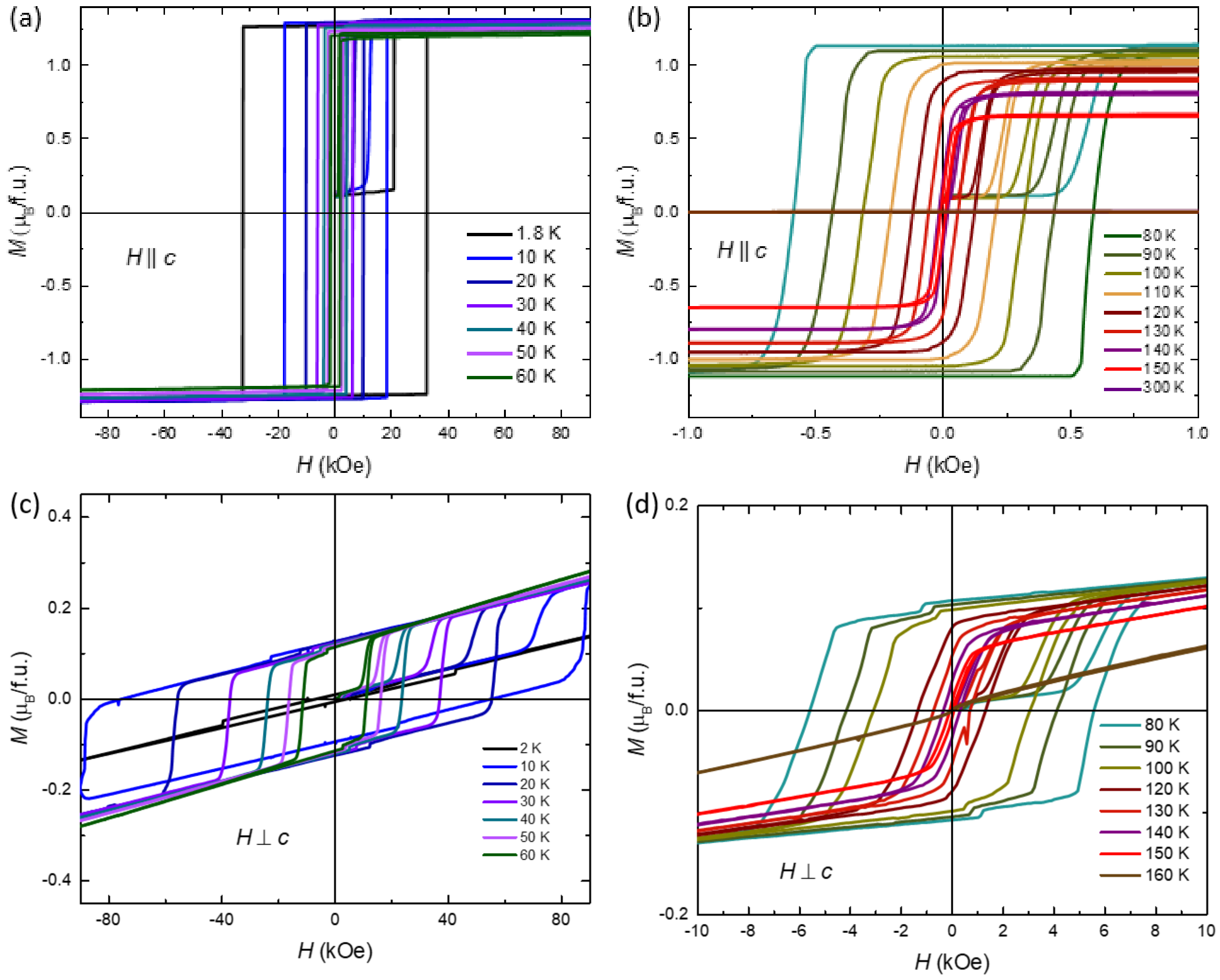

**Fig. 3. Field-dependent magnetization measured at various temperatures along different directions.** Magnetization of a single crystal of $SmCrGe_3$ at different temperatures as a function of the magnetic field applied parallel (**Fig. 3*a***, **Fig. 3*b***) or perpendicular (**Fig. 3*c***, **Fig. 3*d***) to the crystallographic *c*-axis. Each isothermal loop is a 5-quadrant loop. Between loops, the system is taken to 300 K and then cooled in zero field to the next temperature. Since

`

the no-zero remnant field exists, the moment at zero field in the first quadrant is not zero.

**Fig. 3** presents the field-dependent magnetization of a $SmCrGe_3$ single crystal at various temperatures. The low-temperature magnetization loop, when the magnetic field is aligned parallel to the *c*-axis, indicates a hard ferromagnetic material. At a given temperature and with the increasing field, the distribution of domains with different orientations does not change until a certain saturation field is achieved. As shown in **Fig. S3*a***, even with a log scale, the width of the transition to saturation is very small in the low-temperature range. At 1.8 K, the loop exhibits the highest coercivity of approximately 30 kOe. The coercivity decreases with increasing temperature, accompanied by a gradual departure from the loop's initial rectangular configuration at high temperatures. The saturation magnetic moment is around 1.26 $\mu_B$, which is smaller than 1.4 $\mu_B$ in the temperature-dependent magnetization measurements at 40 kOe. The discrepancy in saturation moments may be due to differences in experimental setups. In field-dependent magnetization measurements, the sample was mounted on a straw instead of a quartz rod, which prevented it from falling over a long period in the high magnetic field. When the magnetic field is applied perpendicular to the *c*-axis, the magnetic moment does not reach saturation even at fields up to 90 kOe. Except for the jump features of the magnetization above 1.8 K, the kink features are also observed, forming a small hysteresis at 1.8 K. The reason for the kink features appearing in **Fig. 3*d*** is not known. The larger hysteresis shown above the 1.8 K could be due to the small misalignment perpendicular to the *c*-axis, which should be the projection of easy-axis moments.

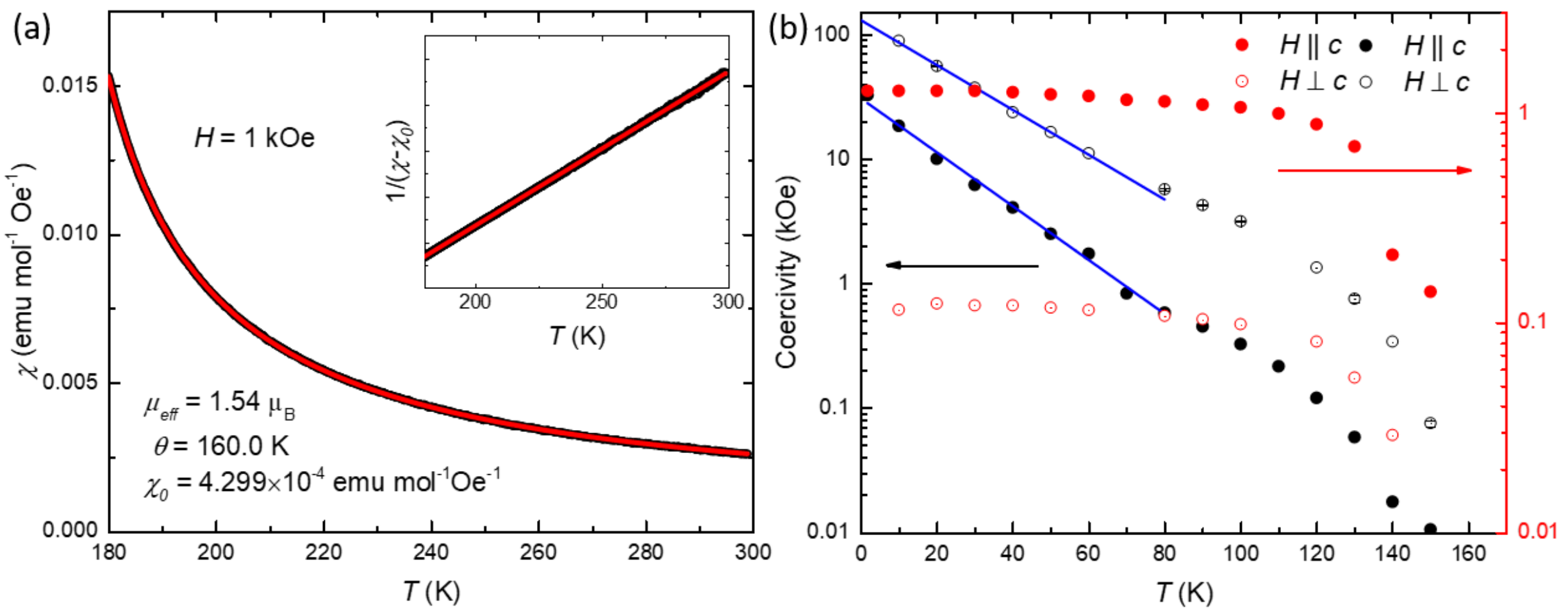

`

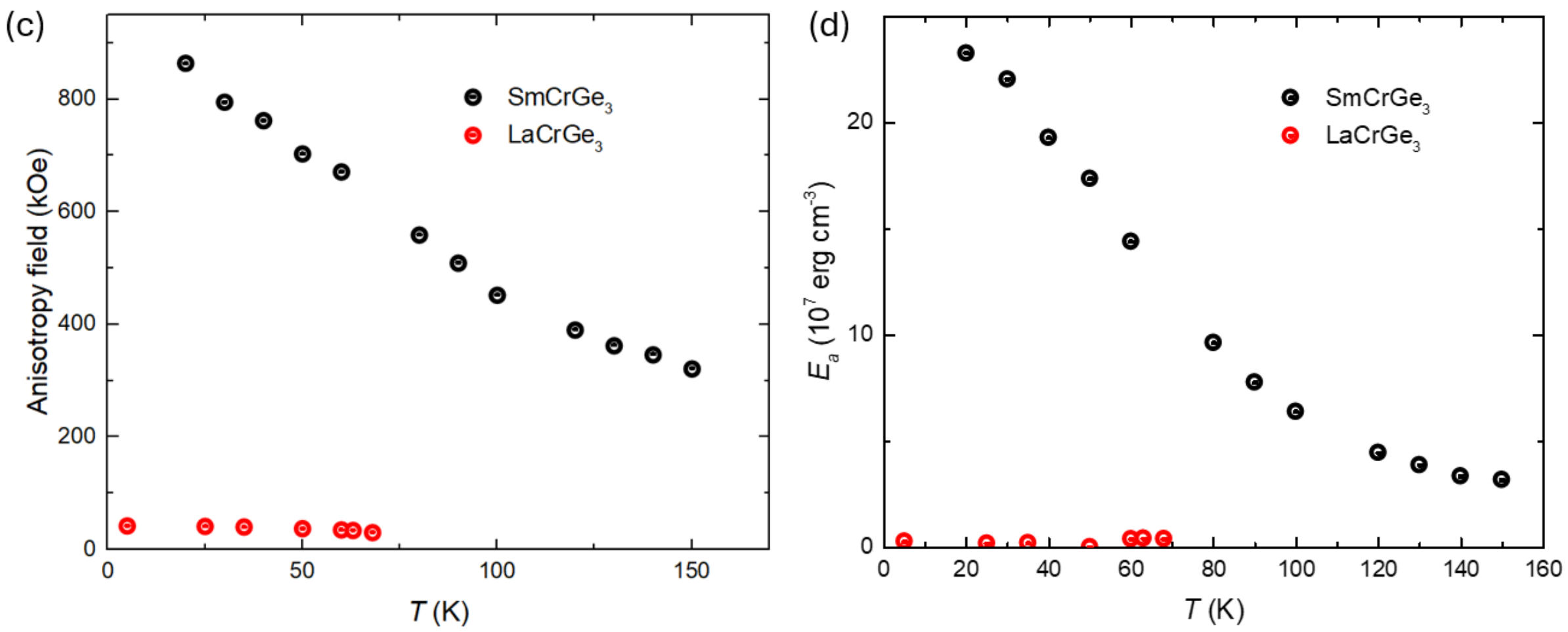

**Fig. 4. Curie-Weiss analysis, coercivity field (black), remanent magnetization (red), anisotropy energy, and M(H) in different $\theta$ (the angle between the *c*-axis and the magnetic field)**. **Fig. 4*a*** shows the Curie–Weiss fitting ($\chi = C/(T - \theta) + \chi_0$) applied on the temperature range from 180 K to 300 K. Inset presents a linear fit to $1/(\chi - \chi_0)$ in a function of temperature. Coercivity field (black) and remanent magnetization (red), $M_R$, as a function of temperature, is shown in **Fig. 4*b*** as the magnetic field parallel (solid) and perpendicular (hollow) to the crystallographic *c*-axis in the log scale. The blue line shows the linear fit on the log scale data. **Fig. 4*c*** shows the anisotropy field of $SmCrGe_3$ (black) and $LaCrGe_3$ (red) as a function of temperature. **Fig. 4*d*** presents the anisotropy energy, $E_a$, of $SmCrGe_3$ (black) and $LaCrGe_3$ (red) as a function of temperature.

The Curie-Weiss model, characterized by the equation $\chi = C/(T - \theta) + \chi_0$, has been employed to fit the magnetization-temperature ($M(T)$) data under an applied magnetic field of 1 kOe for a polycrystalline average data for $SmCrGe_3$, as depicted in **Fig. 4*a***. Here, $\chi$ represents the magnetic susceptibility, C is the Curie constant, $T$ is the temperature, θ is the Weiss temperature, and $\chi_0$ is a temperature-independent susceptibility term. The analysis incorporates a polycrystalline average susceptibility calculated as $\chi = (2\chi_\perp + \chi_\parallel)/3$, where $\chi_\perp$ and $\chi_\parallel$ denote susceptibilities perpendicular and parallel to the *c*-axis, respectively. The fitting yields an effective magnetic moment ($\mu_{eff}$) of approximately 1.54 $\mu_B$ and a Curie-Weiss temperature (θ) of 160.0 K. The θ value is very close to the ferromagnetic transition temperature identified through heat capacity measurements, as shown in **Fig. S4**. Compared with the theoretical $Sm^{3+}$ value of 0.84 $\mu_B$, [36] the fitting effective moment of $SmCrGe_3$ is 1.54 $\mu_B$/formula. Considering results from other $RECrGe_3$ compounds, [14–16, 37] it is difficult to determine if both 3d and 4f electrons contribute to the effective moments. Compared to $LaCrGe_3$, which has an effective moment of 2.5 $\mu_B$/f.u. [12] due to spin

`

fluctuations near the transition in the itinerant system, [38,39] $SmCrGe_3$ likely exhibits more complex contributions to the magnetization tail. These contributions may arise from both the localized moments of 4*f* electrons and itinerant electrons. The decrease of $SmCrGe_3$ effective moment after replacing La with Sm may be due to the potential change of the electronic band structure and the mixed valence of Sm.

The coercivity, a key magnetic property indicative of the resistance to demagnetization, of $SmCrGe_3$ exhibits a pronounced temperature dependence in both orientations relative to the crystallographic axes, as presented in **Fig. 4*b***. Notably, as the linear fitting shown in semi-log the plot, coercivity decreases exponentially as the temperature increases to 80 K, suggesting a change from high to low pinning strength with increasing temperature. **Fig. S3** elucidates these distinctions by comparing the virgin curves of $SmCrGe_3$ and $LaCrGe_3$ single crystals with the applied magnetic field parallel to the *c*-axis. Both materials exhibit rectangular hysteresis loops at lower temperatures, indicating their ferromagnetic nature. Nevertheless, the critical field required to saturate the magnetization of $SmCrGe_3$ surpasses that of $LaCrGe_3$ by several orders of magnitude. Specifically, according to **Fig. S3**, a 0.2 kOe is sufficient to align the magnetic domains in $LaCrGe_3$ as the field along the *c*-axis, whereas $SmCrGe_3$ necessitates a magnetic field magnitude hundreds of times greater to achieve domain alignment. This comparison distinguishes $SmCrGe_3$ as a hard ferromagnet due to magnetocrystalline anisotropy induced by Sm, in contrast to the intrinsically soft ferromagnetic nature of $LaCrGe_3$, highlighting the substantial variation in their magnetic domain behavior and coercivity.

To further study the magnetic anisotropy, the anisotropy field of $SmCrGe_3$ with a comparison to $LaCrGe_3$ as a function of temperature is presented in **Fig. 4*c***, estimated by extrapolating the magnetization-field (M(H)) curves in the high-field regime after jump-like features, with the magnetic field applied perpendicular to the *c*-axis. The evaluation of the magnetic anisotropy within a hexagonal crystal system is quantified by the anisotropy constants, considering the first two anisotropy constants for the simplicity, $K_1$, and $K_2$, according to the expression for anisotropy energy ($E_a$):

$$E_a(\theta) = K_1 sin^2\theta + K_2 sin^4\theta$$

`

And the relation for the anisotropy field (H) per magnetization component perpendicular to the *c*-axis ($M_{ab}$) is given by:

$$\frac{H}{M_{ab}} = \frac{2K_1}{M_s'^2} + \frac{4K_2}{M_s'^4} M_{ab}^2$$

Here, $E_a(\theta)$ represents the anisotropy energy, θ is the angle between the magnetization vector and the *c*-axis, $M_{ab}$ signifies the magnetization component perpendicular to the *c*-axis, and M' denotes the spontaneous magnetization [40]. The analysis employs saturated magnetization for fitting purposes, ensuring that the fitting is conducted in the high-field domain where magnetic domains are predominantly aligned in a singular direction. As demonstrated in **Fig. 4*d***, the anisotropy energy $E_a$ for $SmCrGe_3$ significantly surpasses that of $LaCrGe_3$, indicating a remarkable enhancement in magnetic anisotropy attributed solely to the presence of Sm. This significant disparity underscores the critical role of Sm in enhancing the magnetic anisotropy in $SmCrGe_3$, delineating a stark contrast in the magnetic properties of these two compounds. In linear theory, $K_1 = -3\alpha_J J^2 A_2^0 B_J^2(x) - 40\beta_J J^4 A_4^0 B_J^4(x) - 168\gamma_J J^6 A_6^0 B_J^6(x), K_2 = 35\beta_J J^4 A_4^0 B_J^4(x) + 378\gamma_J J^6 A_6^0 B_J^6(x), x = 2J|g_J - 1|\mu_B B_{ex}/k_B T$, $B_{ex}$ is the exchange field acting on the rare-earth sublattice, $B_J^n(x)$ is the generalized Brillouin function, $A_n^0$ is the uniaxial crystal field parameters and $\alpha_J, \beta_J, \gamma_J$ are the Stevens coefficients. [22,41] According to $K_1$ and $K_2$ expressions, even considering the higher-order and J mixing, [41] 3*d*-4*f* exchange interaction plays an important role in magnetocrystalline anisotropy. This explains the significant increase in anisotropy of $SmCrGe_3$ compared with $LaCrGe_3$.

**Table I: Total magnetic moment M, on-site Sm and Cr magnetic moments, $m_{Sm}$ and $m_{Cr}$, in $SmCrGe_3$ calculated with various *U* values applied on Sm-4f orbitals in DFT+U.** Sm magnetic moment is further resolved into its spin contribution $m_{Sm}^s$ and the contributions from Sm-4*f* spin and orbital, $m_{4f}^s$ and $m_{4f}^l$ . The total magnetization M is in the unit of $\mu_B$/f.u., while all other components are in the unit of $\mu_B$/atom and calculated inside the mun-tin (MT) spheres. The magnetic moment of Ge is negligible (~ 0.02 $\mu_B$/Ge) and not listed. The Cr orbital magnetic moment is negligible compared to its spin moment. The Sm spin moment and Cr magnetic moment have opposite signs, indicating an antiferromagnetic

`

coupling between the Sm and Cr spins. The opposite signs of the Sm 4f orbital and spin magnetic moments reflect the third Hund's rule for the light rare-earth elements. In the large-U limit, the Sm 4*f* total magnetic moment vanishes as its spin and orbital components cancel out, resulting in a small total magnetic moment for Sm ($m_{Sm}$), which is primarily due to the 5*d* spin moments that are parallel to the 4*f* spin moments.

| U | $m_{4f}^{l}$ | $m_{4f}^{s}$ | $m_{Sm}^{s}$ | $m_{Sm}$ | $m_{Cr}$ | $M$ |
|---|---|---|---|---|---|---|
| 6 | 4.17 | -5.18 | -5.34 | -1.17 | 1.58 | 0.35 |
| 8 | 4.53 | -5.11 | -5.25 | -0.72 | 1.54 | 0.74 |
| 10 | 4.81 | -5.00 | -5.18 | -0.37 | 1.51 | 1.05 |
| 12 | 4.92 | -4.97 | -5.16 | -0.24 | 1.51 | 1.18 |
| 14 | 4.95 | -4.97 | -5.16 | -0.21 | 1.52 | 1.21 |

**Table I** summarizes the magnetic moments and their components in $SmCrGe_3$ calculated in DFT+*U* with various *U* values. Here, the signs of the magnetic moments indicate their directions. Note that, according to Hund's rules, $Sm^{3+}$ has a configuration with $S$ = 5/2, $L$ = 5, and $J$ = 5/2. For smaller *U* values, the calculated 4*f* spin and orbital magnetic moments, $m_{4f}^{l}$ and $m_{4f}^{s}$, respectively, show a large deviation from the integer numbers. This is caused by the pining of Sm states at the Fermi level unless a sufficiently large *U* is applied on Sm-4*f* orbitals in DFT+*U*. As a result, the shoulder of the Sm states right above the Fermi level is slightly filled up, resulting in more than five electrons, e.g., ~ 5.18 e at *U* = 6 eV, occupying the Sm-4*f* states. A *U* value larger than 10 eV is required to push the unoccupied 4f states away from $E_F$, ensuring a $Sm^{3+}(f^5)$ configuration. Therefore, in the following, we mainly discuss the calculations performed with *U* = 12 eV.

`

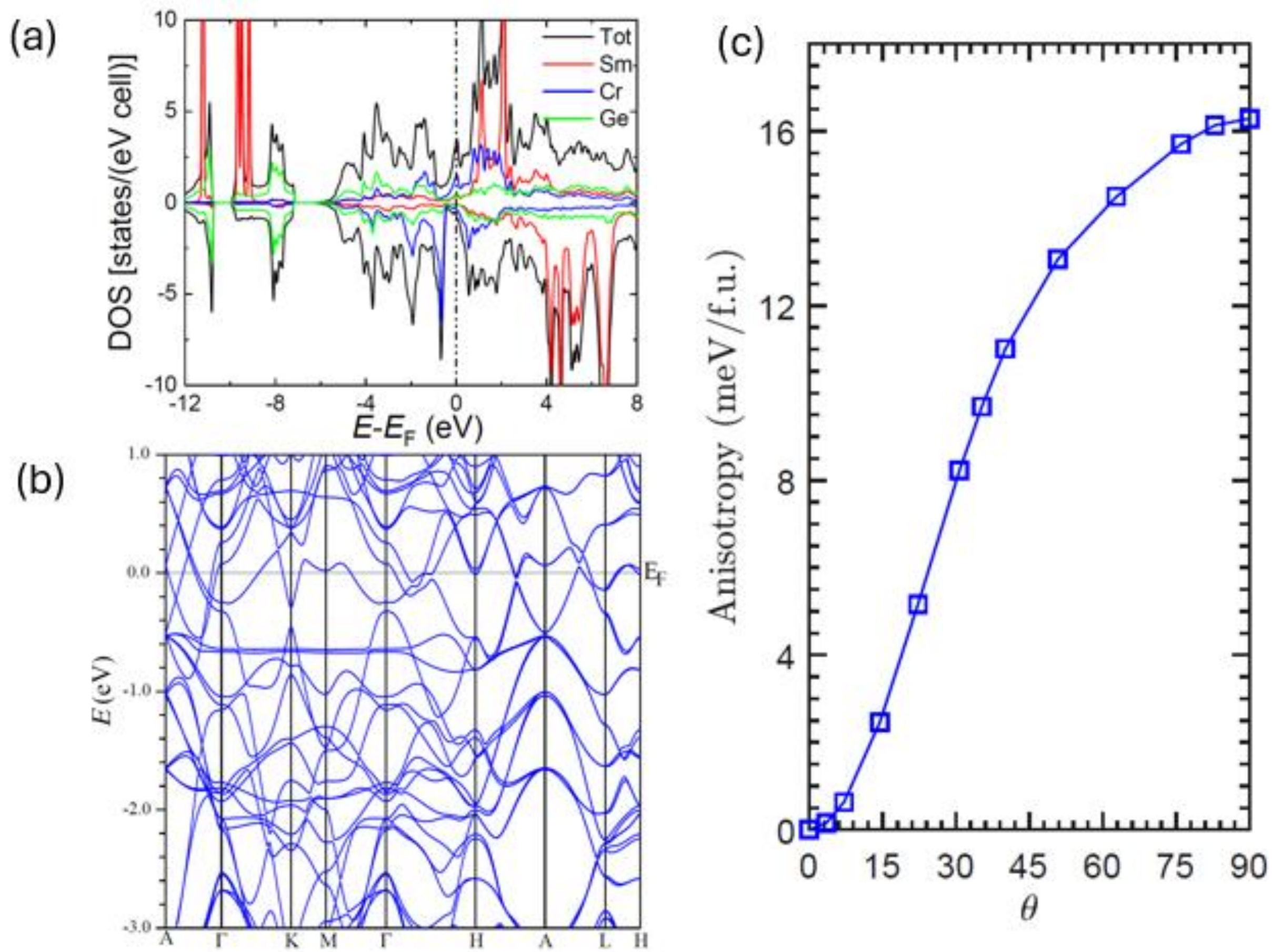

**Fig. 5. DOS, band structure, and magnetocrystalline anisotropy of $SmCrGe_3$, calculated in DFT+U with U = 12 eV applied to Sm-4*f* states**. (***a***) Total and sublattice-decomposed DOS. The total DOS consists of contributions from all atomic sites and the interstitial region. The unit cell contains two formula units. (***b***) Band structure along high symmetry directions. Spin-orbit coupling is included in the calculation. A pair of closely aligned flat bands along the Γ-K-M-Γ is situated at around -0.7 eV below Fermi level. These flat bands become dispersive at finite $k_z$. (***c***) Magnetocrystalline anisotropy, characterized by the variation of magnetic energy (in meV/f.u.) as a function of spin-axis rotation. The spin direction is denoted by the polar angle $\theta$ and the azimuthal angle $\phi$. The lattice vector *c* ([0 0 1]) direction is along the $\hat{z}$ direction and denoted by $\theta = 0°$, while the lattice vector *a* ([1 0 0]) direction is denoted by $\theta = 90°$ and $\phi = 0°$. The calculations are performed with $\phi$ fixed at 60°. A large easy-axis magnetocrystalline anisotropy energy of ~16 meV/f.u. is found.

`

**Fig. 5*a*** shows the total and sublattice-decomposed DOS. At $U$ = 12 eV, the occupied Sm-4*f* states are located at around -10 eV below $E_F$, while the unoccupied majority-spin 4*f* states are in the range of 1-2 eV above $E_F$, with a negligible shoulder at $E_F$, resulting in a $4f^5$ configuration. The occupied states between -6 eV and $E_F$ are mainly Ge-4*p*, Cr-3*d*, and Sm-5*d* states, with expected hybridization between them. The Ge-4*s* states are present around the -12 to -8 eV region. The Cr-3*d* minority-spin channel shows a peak at $E_F$, and its majority-spin channel has a larger sharp peak at ~ 0.7 eV below $E_F$, attributed to a flat band as discussed below.

**Fig. 5*b*** shows the electronic band structure of $SmCrGe_3$ calculated along high symmetry directions, primarily in the $k_z = 0$ and 0.5 (2π/c) planes. Noticeably, two seemingly parallel flat bands are around -0.7 eV in the Γ-K-M-Γ direction. These two bands are dominated by Cr-$d_{xy}$ and Cr-$d_{xz}$ characters. The seemingly universal small splitting between them along Γ-K-M-Γ is actually caused by the combination of crystal field and SOC. In the absence of SOC, these two bands degenerate at high symmetric *k* points and slightly split elsewhere. SOC further lifts the band degeneracy, especially by introducing the largest splittings at these otherwise-degenerate high symmetric *k*-points, resulting in overall two seemingly parallel flat bands. It is worth noting that a previous study has reported flat bands at -0.15 eV below $E_F$ and has attributed it to be the source of the magnetic fragility of $LaCrGe_3$ [22]. However, the flat bands in $SmCrGe_3$ we found here are much farther away from $E_F$, unlikely to play a similar role as reported in $LaCrGe_3$. On the other hand, unlike La, Sm-4*f* spin can facilitate Cr spin polarization via the exchange coupling mediated by Sm-5*d* electrons [42].

**Fig. 5*c*** shows the calculated total energies $E(\theta)$ as functions of spin-quantization direction, characterized by the polar angle $\theta$ and the azimuthal angle $\phi$. The spin axis rotates from the [001] direction to the [110] direction, showing uniaxial anisotropy. The energy minimum occurs at [001], suggesting easy-axis anisotropy in $SmCrGe_3$. The calculated uniaxial anisotropy of $K_U$ =16 meV/f.u. is in reasonable agreement with the experimental value of 13 meV/f.u.(measured at 20 K and $K_U = K_1 + K_2$) estimated from

`

magnetization measurements. Moreover, the out-of-plane energy profile curve deviates from a sinusoidal shape, suggesting the significance of higher-order crystal parameters in contributing to the magnetic anisotropy [42]

## Conclusions

Hexagonal $SmCrGe_3$ is characterized by significant magnetocrystalline anisotropy, with density functional theory (DFT) calculations revealing an enhancement of magnetocrystalline anisotropy of about 16 meV/f.u. at 0 K. This value closely aligns with experimental observations, which document an anisotropy of 13 meV/formula unit (f.u.) at 20 K, comparable to that of $SmCo_5$ (13 – 15 meV/f.u. at base temperature [43]). The synthesis of $SmCrGe_3$ single crystals is achieved through the flux growth method, with subsequent magnetic and specific heat measurements delineating a ferromagnetic transition temperature of approximately 160 K compared with a $T_C$ of 155 K polycrystalline sample. [11] These studies highlight the influence of 4*f* and 3*d* electrons on the critical contribution towards the noted magnetic anisotropy due to the exchange interaction and crystal field.

## Acknowledgments

The work at Michigan State University was supported by the U.S.DOE-BES under Contract DE-SC0023648. X.K. acknowledges the financial support by the U.S. Department of Energy, Office of Science, Office of Basic Energy Sciences, Materials Sciences and Engineering Division under DE- SC0019259. Work in Ames was supported by the U.S. Department of Energy, Office of Science, Basic Energy Sciences, Materials Sciences and Engineering Division under Contract No. DE-AC02-07CH11358.

`

`

# Appendix

**Table S1.** The crystal structure and refinement of $SmCrGe_3$ at room temperature K (Mo Kα radiation). Values in parentheses are estimated standard deviation from refinement.

| **Chemical Formula** | **$SmCr_{0.89}Ge_3$** |
|---|---|
| Formula Weight | 415.18 g/mol |
| Space Group | $P6_3/mmc$ |
| Unit Cell dimensions | $a$ = 6.0898(3) Å<br>$b$ = 6.0898(3) Å<br>$c$ = 5.6666(3) Å |
| Volume | 181.99(2) Å$^3$ |
| Z | 2 |
| Density (calculated) | 7.576 g/cm$^3$ |
| Absorption coefficient | 43.109 mm$^{-1}$ |
| F (000) | 359 |
| 2θ range | 7.728 to 82.192° |
| Reflections collected | 5510 |
| Independent reflections | 262 [$R_{int}$ = 0.0764] |
| Refinement method | Full-matrix least-squares on F$^2$ |
| Data/restraints/parameters | 262/0/11 |
| Final $R$ indices | $R_1$ (I>2σ(I)) = 0.0253; $wR_2$ (I > 2 σ(I)) = 0.0662<br>$R_1$ (all) = 0.0260; $wR_2$ (all) = 0.0672 |
| Largest diff. peak and hole | +3.87 e$^-$/Å$^3$ and -2.60 e$^-$/Å$^3$ |
| R. M. S. deviation from mean | 0.402 e$^-$/Å$^3$ |
| Goodness-of-fit on F$^2$ | 1.115 |

**Table S2.** Atomic coordinates and equivalent isotropic atomic displacement parameters (Å$^2$) of $SmCrGe_3$. ($U_{eq}$ is defined as one-third of the trace of the orthogonalized $U_{ij}$ tensor.)

| **$SmCr_{0.89}Ge_3$** | **Wyck.** | ***x*** | ***y*** | ***z*** | **Occ.** | **$U_{eq}$** |
|---|---|---|---|---|---|---|
| **Sm** | 2*d* | 1/3 | 2/3 | 3/4 | 1 | 0.006(1) |
| **Ge** | 6*h* | 0.1931(1) | 0.3862(1) | 1/4 | 1 | 0.005(1) |
| **Cr** | 2*a* | 0 | 0 | 0 | 0.906(9) | 0.005(1) |

`

**Tables S1 and S2** show the results of the single-crystal XRD. The structure was solved and refined using the Bruker SHELXTL Software Package with the space group *P6₃/mmc*, $SmCr_{0.906(7)}Ge_3$. The final anisotropic full-matrix least-squares refinement on $F^2$ with 11 variables converged at $R_1$ = 2.60 %, for the observed data and $wR_2$ = 6.72 % for all data. The goodness-of-fit was 1.115. The largest peak in the final difference electron density synthesis was 3.87 $e^-/Å^3$, and the largest hole was -2.60 $e^-/Å^3$ with an RMS deviation of 0.402 $e^-/Å^3$. Based on the final model, the calculated density was 7.576 $g/cm^3$ and F (000), 359 $e^-$.

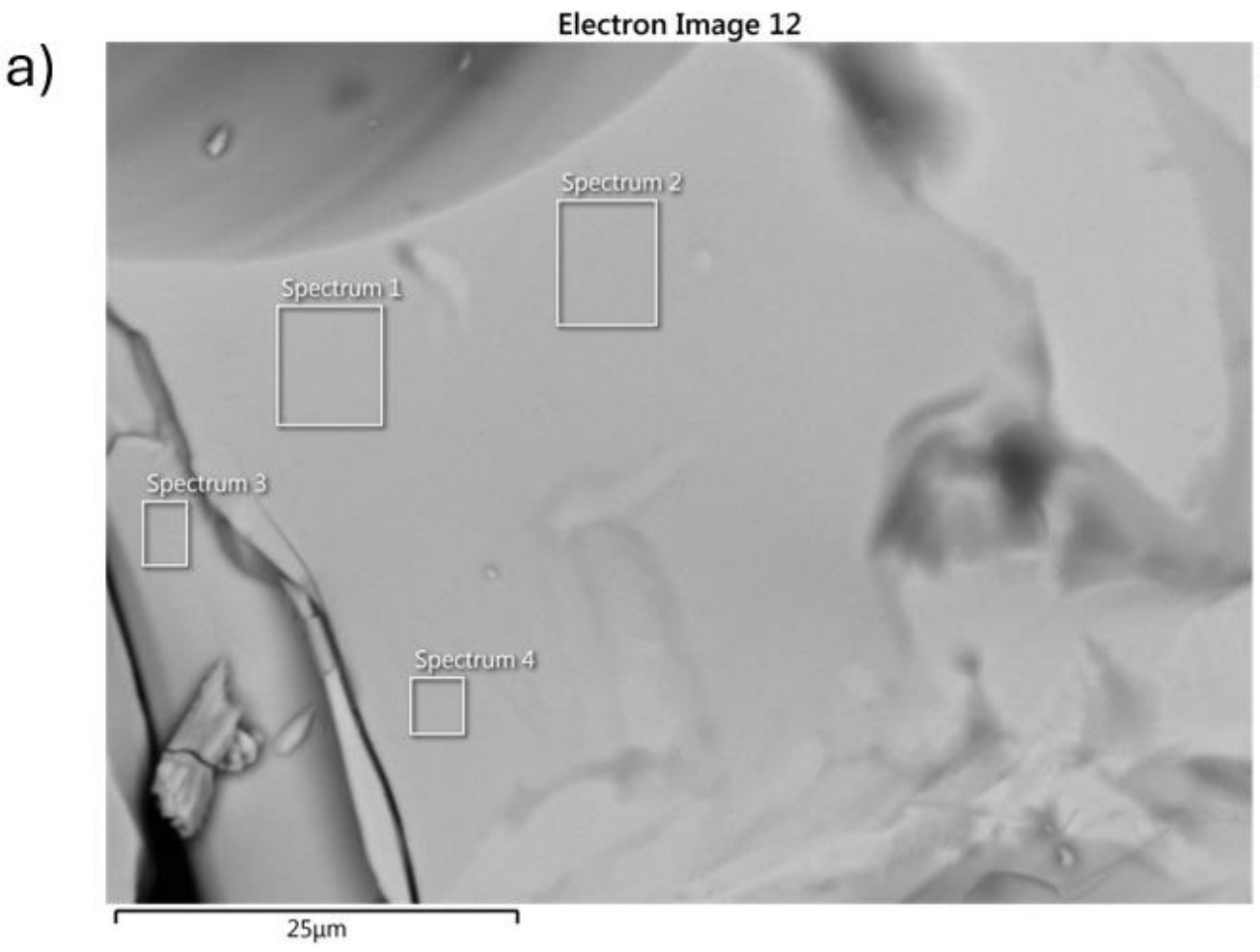

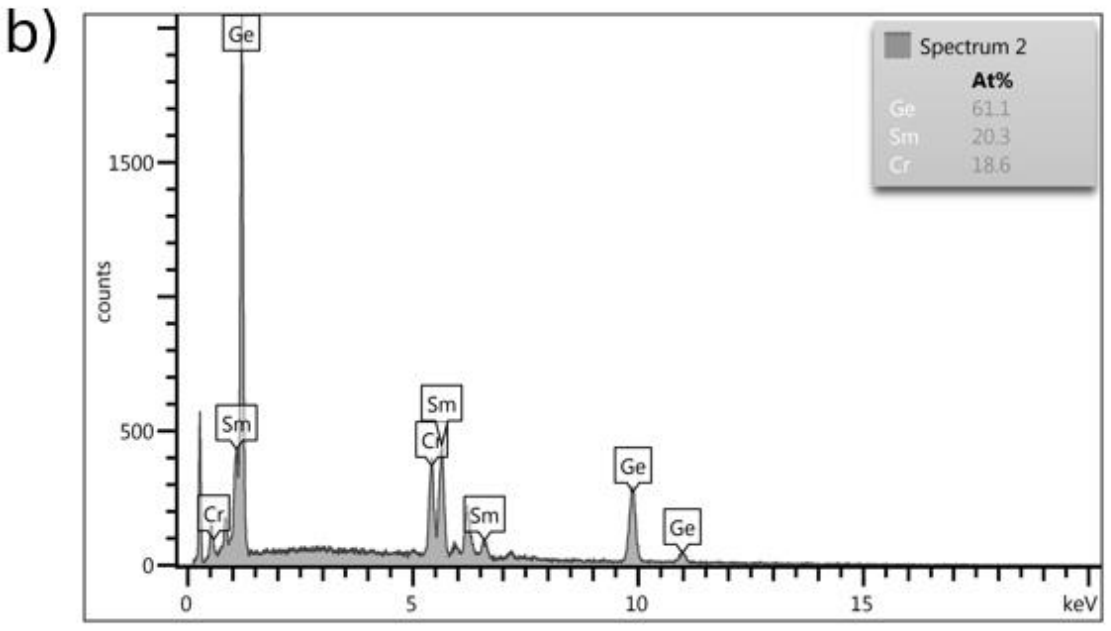

c)

| | |
|---|---|
| Sm | 1 |
| Cr | 0.93(2) |
| Ge | 2.97(7) |

**Fig. S1. SEM and EDS results of single crystals SmCrGe3**. **Fig. S1*a*** gives the SEM image of one of the measured samples. **Fig. S1*b*** shows one of the EDS spectrum results. **Fig. S1*c*** presents the statistical results of 13 spectra from two different pieces of samples.

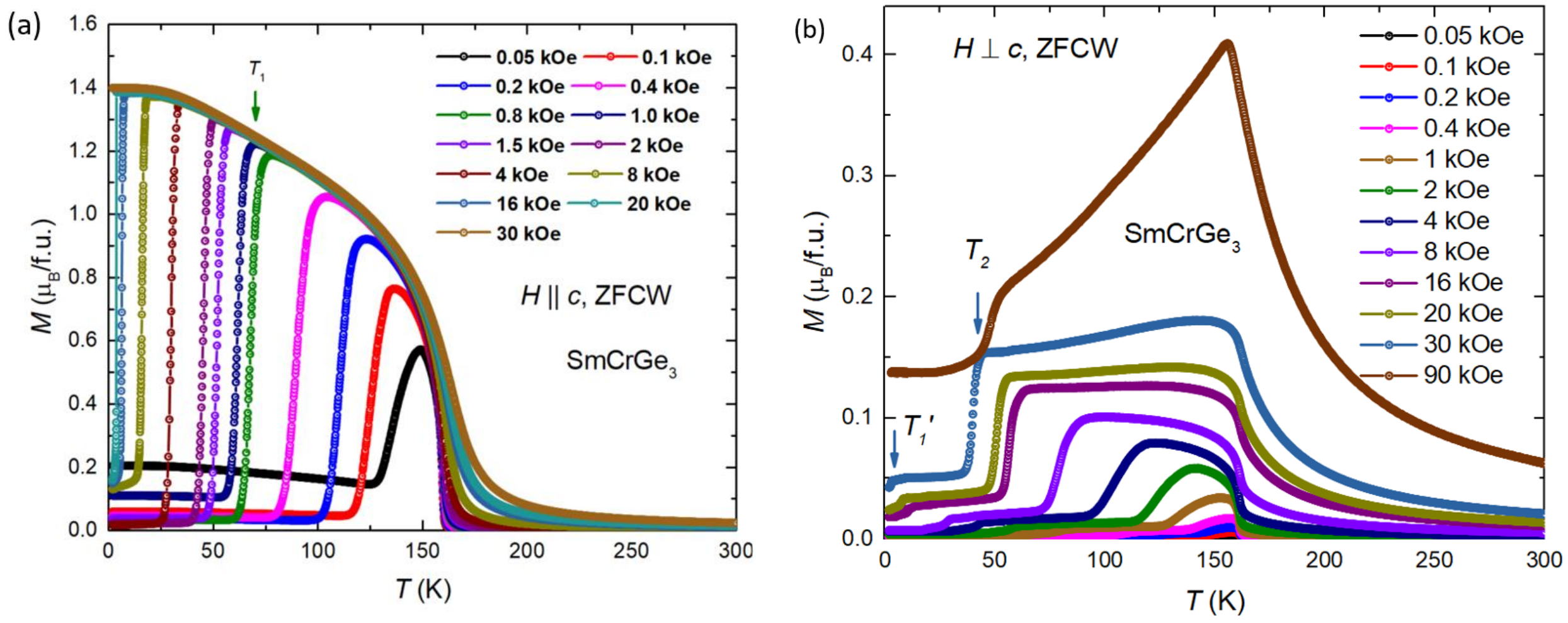

`

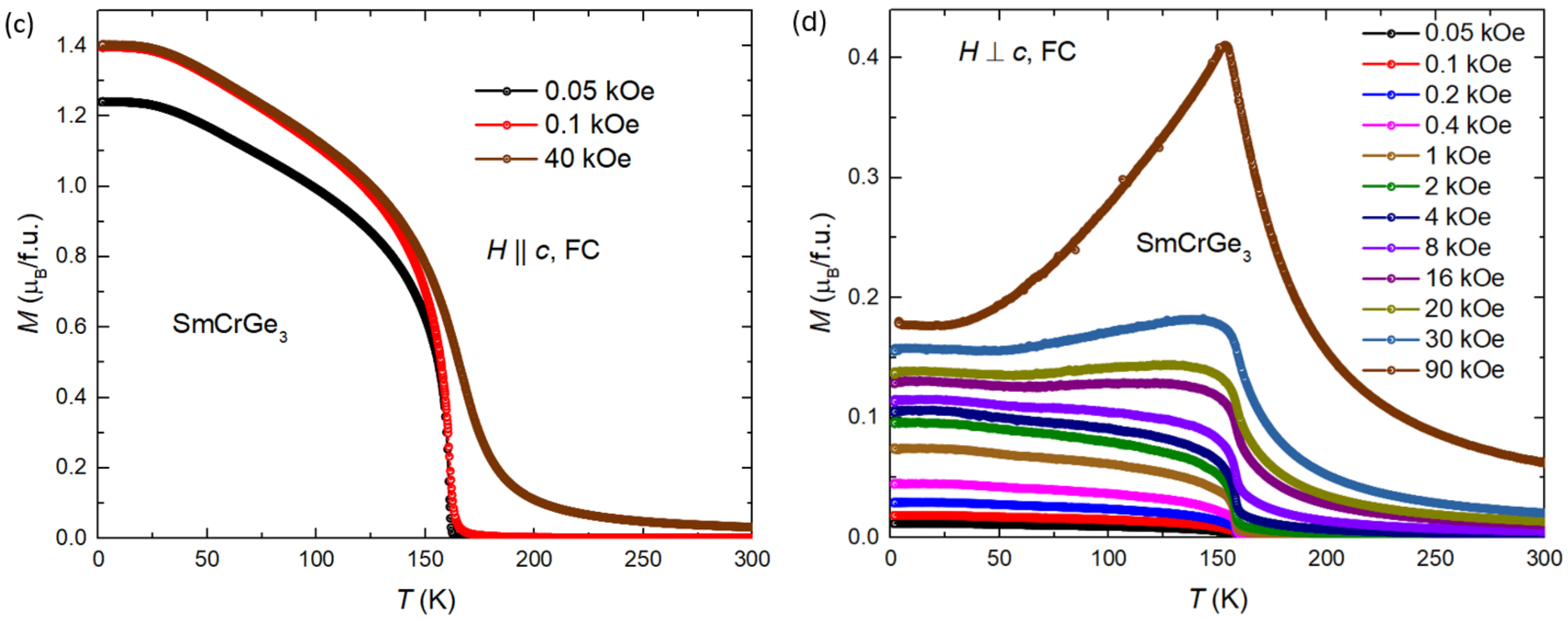

**Fig. S2. Temperature-dependent magnetization of $SmCrGe_3$ measured along different directions at various applied fields in zero-field-cooled-warming (ZFCW) and field-cooled-warming mode.** (**Fig. S2*a***) measurements conducted along the *c*-axis in ZFCW. $T_1$ is the feature temperature that characterizes the jump-like feature as the field is 0.8 kOe. (**Fig. S2*b***) measurements are conducted perpendicular to the *c*-axis in ZFCW. $T_1$' and $T_2$ are the feature temperatures that characterize the kink-like features as the field is 30 kOe. (**Fig. S2*c***) measurements conducted along the *c*-axis in FC. (**Fig. S2*d***) measurements conducted perpendicular to the *c*-axis in FC.

**Fig. 2S*a*** and **2S*c*** show the $M(T)$ under different magnetic fields parallel to the *c*-axis under the ZFCW and FC protocol. As the field increases, the jump-like features in ZFCW measurements, denoted as $T_1$, are suppressed as the magnetic field increases. When the magnetic field increases to 30 kOe, the jump-like features disappear, and there is no hysteresis. **Fig. 2S*b*** shows the temperature-dependent magnetization with a field perpendicular to the *c*-axis under the ZFCW temperature protocol. As shown in the 30 kOe magnetization data, two kink-like features are shown and denoted as $T_1$' and $T_2$. Both of these feature temperatures decrease as the magnetic field increases. As the magnetic field is smaller than 8 kOe, the magnetization increases after transition, then decreases around $T_2$. When the magnetic field is larger than 16 kOe, the magnetization directly decreases after the transition. This drop of magnetization before $T_2$ becomes larger as the magnetic field increases. When the magnetic field reaches 90 kOe, this decreased value becomes significant and reaches almost half the maximum magnetization near the transition temperature. **Fig. 2S*c*** gives magnetization as a function of temperature in the magnetic field parallel to the *c*-axis under FC. Below 0.1 kOe, the magnetization at the base temperature increases as the magnetic field increases. After 0.1 kOe, there is no observable magnetization change at base temperature as the field increases.

`

**Fig. 2S*d*** shows the temperature-dependent magnetization in the magnetic field up to 90 kOe perpendicular to the *c*-axis under FC. When the magnetic field is below 8 kOe, the magnetization increases as the temperature decreases and reaches maximum at the base temperature. When the applied field is larger than 16 kOe, after transition, the magnetization decreases first, then increases a little. As the magnetic field goes up to 90 kOe, the magnetization significantly decreases as the temperature decreases. In **Fig. 2S*a***, the magnetization value reaches around 1.4 $\mu_B$ at 30 kOe when the field is parallel to the *c*-axis; however, when the field is perpendicular to the *c*-axis, even the magnetic field is three times larger than 30 kOe, the maximum magnetization is only less than half of the magnetization in another direction. This indicates the magnetocrystalline anisotropy in $SmCrGe_3$.

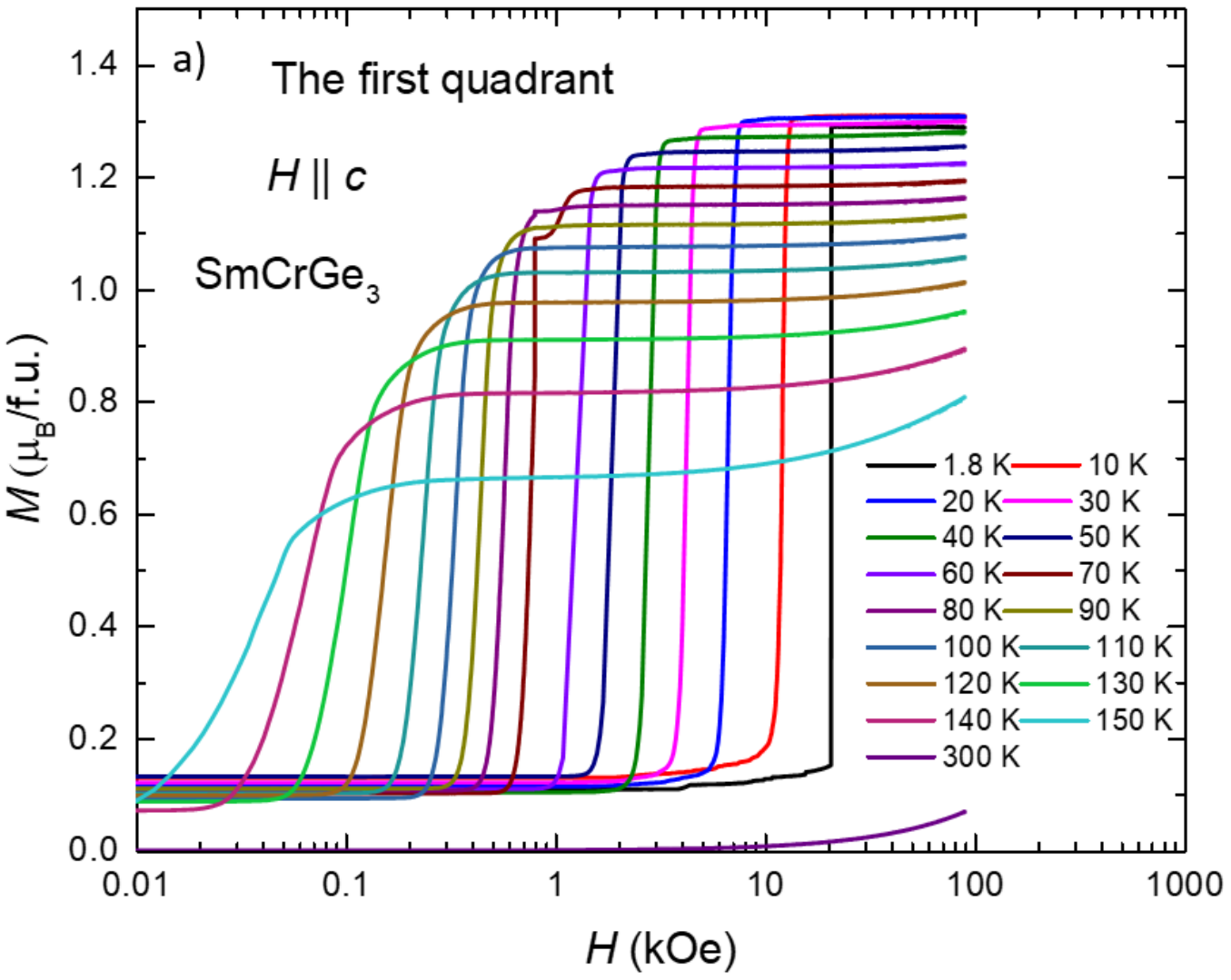

`

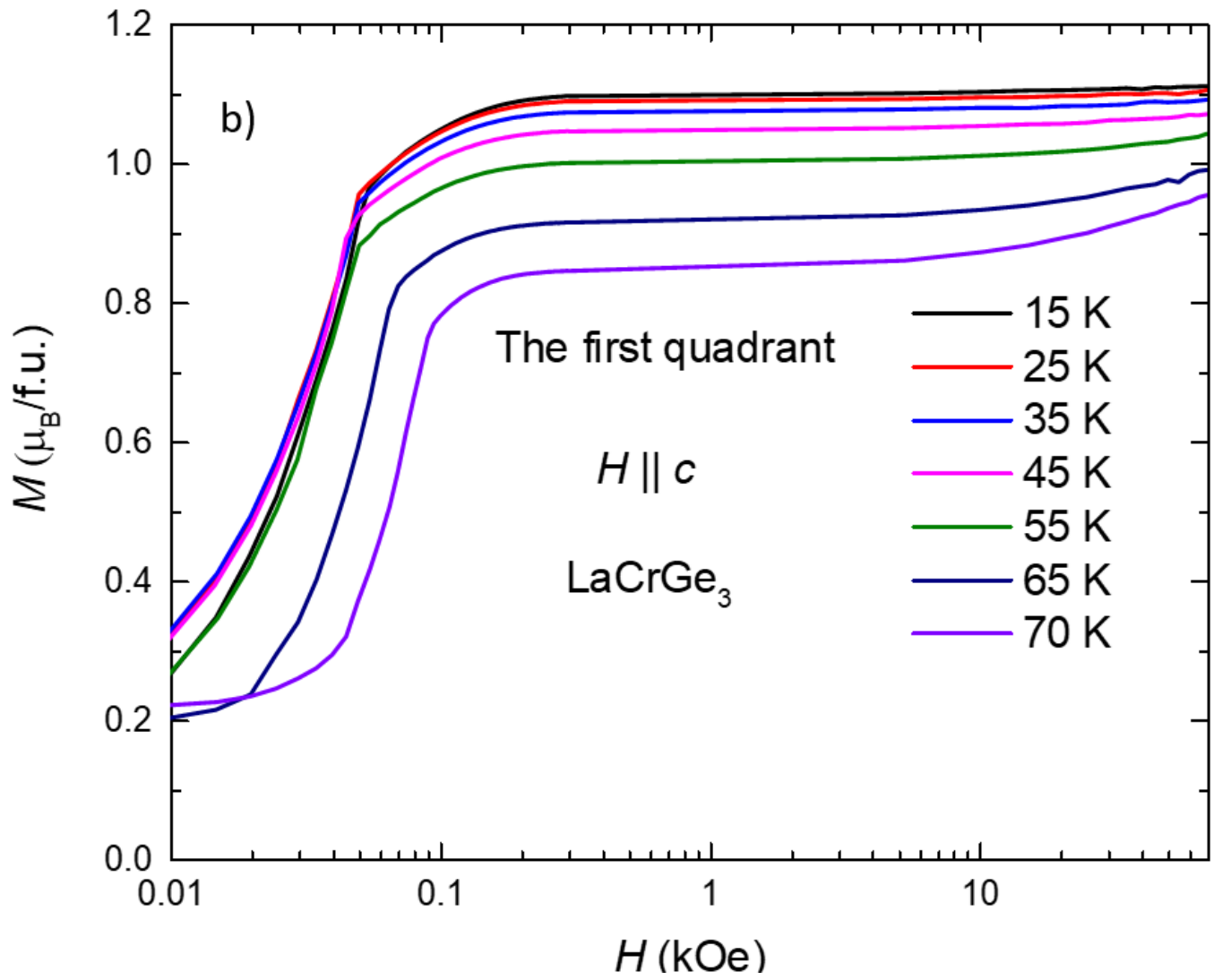

**Figure S3. First-quadrant field-dependent magnetization**. The first-quadrant field-dependent magnetization is plotted for both $SmCrGe_3$ and $LaCrGe_3$ single crystal as a field parallel to the *c*-axis.

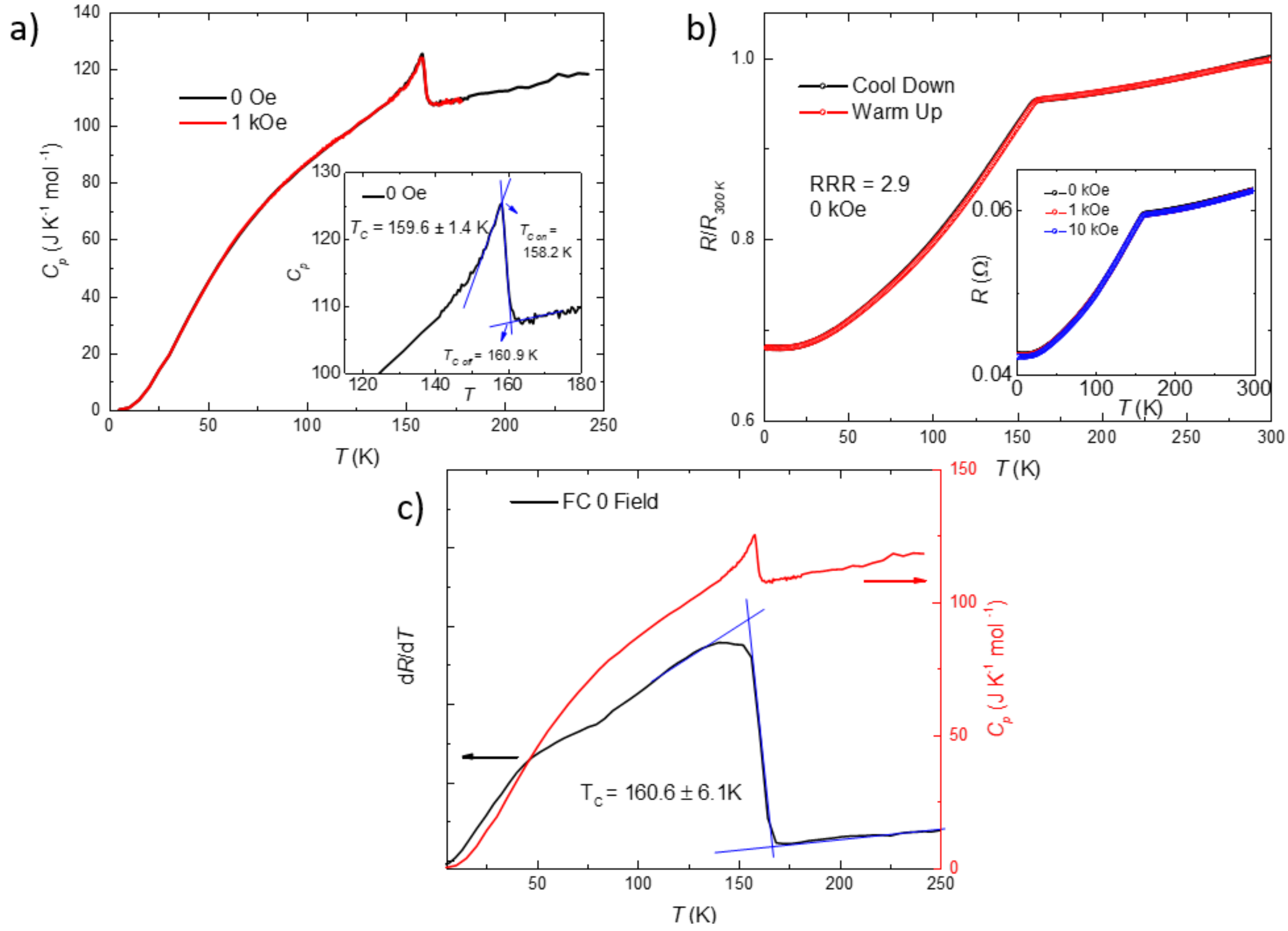

**Fig. S4. Temperature-dependent specific heat. Fig. S4*a*** shows the specific heat as a function of temperature under 0 (black) and 1 kOe (red). The (Insert) criteria used to determine $T_C$. **Fig. S4*b*** shows temperature-dependent resistance, and insert gives the

`

measurements under different fields. **Fig. S4*c*** presents the comparison between d*R*/d*T* and $C_p(T)$.

The ferromagnetic transition is determined by temperature-dependent specific heat and resistance measurements, as shown in **Fig. S4a** and **b**. Measurements are taken under 0 kOe and 1 kOe field perpendicular to the *c*-axis. These two measurements overlap, indicating negligible low-field influence on the specific heat capacity. A distinct second-order phase transition is observed, characterized by a well-defined peak in the specific heat curve, pinpointing the ferromagnetic transition temperature at 159.6 ± 1.4 K. In **Fig. S4*b***, the clear kink in temperature-dependent resistance is observed. The magnetic field, up to 10 kOe, was applied along *ab* directions, with the current flowing in the *c* direction. There is no clear different between zero and 10 kOe temperature-dependant resistance measurements. The transition temperature from resistance measurement is 160.6 ± 6 K, according to **Fig. S4*c***, which is close to the result of specific heat measurements with a larger transition width.

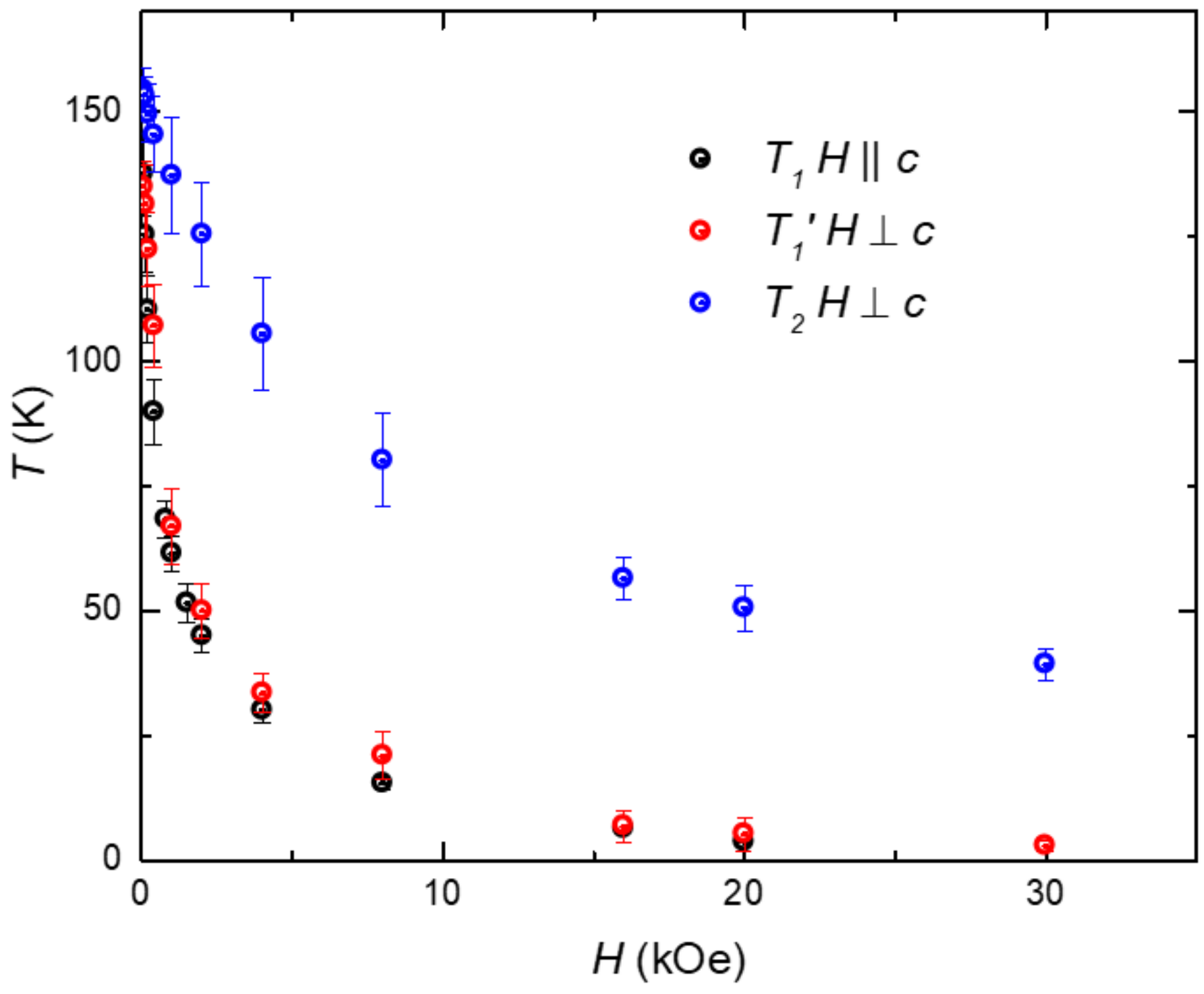

**Fig. S5 diagram of $T_1$, $T_1'$ and $T_2$. Fig. S5** shows jump temperature as a function of the magnetic field as the field parallel (black) and perpendicular (red and blue) to the crystallographic *c*-axis.

`

The temperatures associated with these jumps and kink-like features in zero-field-cooled-warming (ZFCW) magnetization are presented in **Fig. S5** as $T_1$, $T_1'$, and $T_2$. These temperates are shown in **Fig. 2*a*** and **2*b***. An overlap between $T_1$ and $T_1'$, the minimal change in magnetization of $T_1'$, suggests that $T_1$ and $T_1'$ correspond to identical magnetic features. The reason $T_1'$ appears in the field perpendicular to the *c*-axis is not known. Considering the demagnetization field, $T_1$ and $T_2$ may change due to the shape of the different samples. This will not be discussed in this paper. Moreover, the analysis reveals a decrease in $T_1$ and $T_2$ values with increasing magnetic field intensity, further elucidating the magnetic characteristics of $SmCrGe_3$.